\documentclass[aps,prb,twocolumn,longbibliography]{revtex4-2}
\usepackage{amsfonts}
\usepackage{mathrsfs}
\usepackage{amsmath}
\usepackage{color}
\usepackage{natbib}
\usepackage{graphicx}
\usepackage{bm}
\usepackage{amssymb}
\usepackage{xspace}
\usepackage{epstopdf}
\usepackage{dcolumn}
\usepackage{longtable}
\usepackage{multirow}
\usepackage[colorlinks=true, letterpaper=true, pdfstartview=FitV, linkcolor=blue, citecolor=blue, urlcolor=blue]{hyperref}
\usepackage{gensymb}
\usepackage{gensymb}

\makeatletter

\newcommand{\Rmnum}[1]{\expandafter\@slowromancap\romannumeral #1@}
\makeatother

\begin{document}
	
	\title{Engineering Ideal 2D Type-II Nodal Line Semimetals via Stacking and Intercalation of van der Waals Layers}
	
\author{Li Chen}
\affiliation{College of Physics and Electronic Engineering, Center for Computational Sciences, Sichuan Normal University, Chengdu, 610068, P.R.China}
\author{Junlan Shi}
\thanks{These authors contributed equally to this work.}
\affiliation{College of Physics and Electronic Engineering, Center for Computational Sciences, Sichuan Normal University, Chengdu, 610068, P.R.China}
 \author{Jiani Zhang}
\affiliation{College of Physics and Electronic Engineering, Center for Computational Sciences, Sichuan Normal University, Chengdu, 610068, P.R.China}
\author{Botao Fu}
\email[]{fubotao2008@gmail.com}
\affiliation{College of Physics and Electronic Engineering, Center for Computational Sciences, Sichuan Normal University, Chengdu, 610068, P.R.China}

	\date{\today}

\begin{abstract}
Two-dimensional type-II topological semimetals (TSMs), characterized by strongly tilted Dirac cones, have attracted intense interest for their unconventional electronic properties and exotic transport behaviors. However, rational design remains challenging due to the sensitivity of band tilting to lattice geometry, atomic coordination, and symmetry constraints. Here, we present a bottom-up approach to engineer ideal type-II nodal line semimetals (NLSMs) in van der Waals bilayers via atomic intercalation. Using monolayer $h$-AlN as a prototype, we show that fluorine-intercalated bilayer AlN (F@BL-AlN) hosts a symmetry-protected type-II nodal loop precisely at the Fermi level, enabled by preserved mirror symmetry ($\mathcal{M}_z$) and tailored interlayer hybridization. First-principles calculations reveal that fluorine not only tunes interlayer coupling but also aligns the Fermi energy with the nodal line, stabilizing the type-II NLSM phase. The system exhibits tunable electronic properties under external electric and strain fields and features a van Hove singularity that induces spontaneous ferromagnetism, realizing a ferromagnetic topological semimetal state. This work provides a versatile platform for designing type-II NLSMs and offers practical guidance for their experimental realization.
			
\end{abstract}
	
	\maketitle

\section{Introduction}

Two-dimensional topological semimetals provide an exceptional platform for exploring unconventional electronic states arising from symmetry-protected band crossings~\cite{armitage2018weyl,Weng2016,yu2022encyclopedia}. Among them, nodal-line semimetals (NLSMs)~\cite{yu2017topological,yang2018symmetry,PhysRevB155156}, characterized by one-dimensional manifolds of band degeneracy in momentum space, host a range of intriguing phenomena including nontrivial Berry phases, drumhead-like boundary states, and anomalous transport responses~\cite{zhao2022two,he2018type,PhysRevB.107.115168}. While numerous NLSMs have been theoretically proposed~\cite{PhysRevB.96.081106,PhysRevB.93.235147,zhang2017topological,PhysRevB.103.125425,PhysRevResearch.1.032006} and experimentally identified, the realization of type-II NLSMs—defined by strongly overtilted band dispersions and a finite density of states at the nodal energy—remains particularly elusive in two dimensions~\cite{chang2019realization,zhao2024observation,PhysRevB.105.115142}.

The fundamental difficulty originates from the fact that type-II nodal lines (NLs) require a delicate and highly nontrivial balance between band inversion and momentum-dependent band tilting. Both ingredients are exquisitely sensitive to lattice geometry, orbital hybridization, and symmetry constraints\cite{PhysRevB.95.094513,rosenstein2023two,PhysRevB.103.L081402,li2025type}. In realistic solid materials, these factors are often strongly entangled, making them difficult to tune independently. As a consequence, although symmetry indicators and high-throughput screening have greatly accelerated the discovery of topological semimetals in general~\cite{PhysRevX.8.031069,zhang2021predicting,PhysRevB.109.205141,bradlyn2016beyond}, they are largely ineffective for predicting or stabilizing 2D type-II NL phases, whose existence hinges on fine electronic-structure details beyond conventional descriptors~\cite{gao2021high,PhysRevMaterials.8.024201}.

\begin{figure}
	\includegraphics[width=3.3 in]{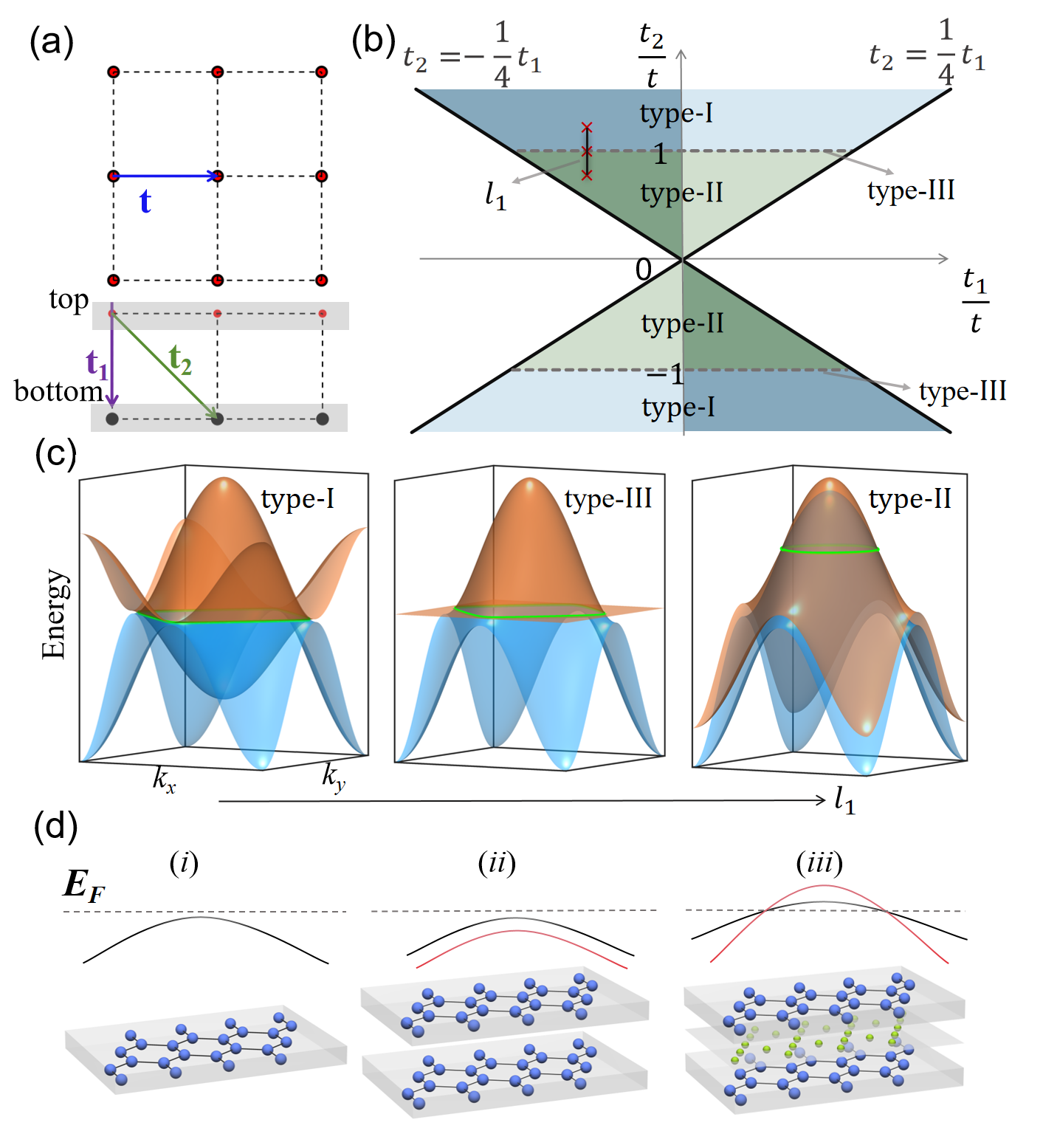}
	\caption{(a) Schematic illustration of the bilayer square-lattice model with intralayer hopping $t$ and interlayer hoppings $t_{1}$ and $t_{2}$. (b) Phase diagram of the minimal tight-binding model in the $(t_{1}/t, t_{2}/t)$ parameter space, showing three distinct NL phases. (c) Representative 3D band dispersions corresponding to the three NL types along the path $l_{1}$ indicated in (b). (d) Schematic illustration of the three-step bottom-up strategy for designing ideal 2D type-II NLSMs using van der Waals bilayer stacking and ionic intercalation.}\label{Fig1}
\end{figure}

This challenge naturally calls for a bottom-up materials design strategy that enables precise and independent control over interlayer coupling, symmetry protection, and Fermi-level positioning. In this work, we demonstrate that van der Waals stacking combined with atomic intercalation provide exactly such a platform. By stacking two identical monolayers to preserve mirror symmetry and subsequently introducing guest atoms into the van der Waals gap, one can selectively enhance interlayer hybridization while simultaneously tuning the carrier density. This dual functionality offers an experimentally viable route to engineer overtilted nodal dispersions and to align the nodal energy with the Fermi level.

Guided by these principles, we construct a minimal tight-binding model that reveals the competition between intra- and interlayer hopping as the key mechanism for realizing type-II nodal-line (NL) dispersions. Translating this insight into a concrete materials platform, we focus on hexagonal nitrides ($h$-XN, X = B, Al, Ga) and identify fluorine-intercalated bilayer AlN (F@BL-AlN) as an ideal two-dimensional type-II NLSM, hosting a mirror-symmetry-protected nodal loop precisely pinned at the Fermi level.
Crucially, this type-II NLSM is not merely a static topological phase but a highly tunable platform. External electric fields and biaxial strain allow efficient and reversible control over the nodal topology and its symmetry protection. Furthermore, the strongly overtilted dispersion generates a finite density of states at the nodal energy, enhancing electronic correlations and naturally promoting magnetic instabilities, thus paving the way for magnetic topological semimetals.
Our results provide clear physical insight and general design principles for realizing and manipulating 2D type-II NLSMs, offering a versatile and experimentally accessible platform to explore the interplay between tunable topology, electronic correlations, and magnetism.

\section{A Minimal Model for Type-II Nodal Line Semimetals}
To elucidate the fundamental physics governing the emergence of type-II nodal line semimetals (NLSMs), we develop a minimal two-orbital tight-binding (TB) model defined on a bilayer square lattice with mirror symmetry. A corresponding model on a triangular lattice is provided in Fig.~S1 of the Supplementary Material for comparison. As depicted in Fig. 1(a), this model incorporates intralayer nearest-neighbor hopping with amplitude \( t \), along with two distinct types of interlayer coupling: a vertical hopping \( t_1 \) and a skew hopping \( t_2 \). The resulting momentum-space Hamiltonian takes the form:
\begin{equation}\label{ham1}
\begin{split}
H(\mathbf{k}) &= 2t\big[\cos(k_x)+\cos(k_y)\big]\sigma_0 \\
&\quad + \big[t_1 + 2t_2(\cos(k_x)+\cos(k_y)) \big]\sigma_x,
\end{split}
\end{equation}
where \( \sigma_0 \) and \( \sigma_x \) denote the identity and Pauli matrices acting in the layer subspace, respectively. Despite its elegant minimalism, this model successfully captures the essential competition between intra- and interlayer hopping processes that dictates the formation and dispersion of NLs.

A central finding of our analysis is that the existence of NLs is primarily controlled by the ratio of the two interlayer hoppings. As summarized in the phase diagram of Fig.~\ref{Fig1}(b), NLs emerge robustly in the regime where \( |t_2| > \frac{1}{4}|t_1| \) (shaded region). More intriguingly, the type of nodal dispersion—whether type-I, type-II, or type-III—is determined by the interplay between interlayer coupling \( t_2 \) and intralayer hopping \( t \).
Tracing the path \( l_1 \) in parameter space reveals a continuous topological evolution: when \( |t_2/t| > 1 \), interlayer coupling dominates, yielding a conventional type-I NL. As \( |t_2/t| \) decreases below unity, however, enhanced intralayer hopping induces a pronounced band tilting that overturns the conventional quasiparticle dispersion, stabilizing a type-II NL. Exactly at the Lifshitz transition \( |t_2/t| = 1 \), the band flattens completely, resulting in a rare type-III NL. The smooth evolution across these distinct topological regimes is summarized in Fig.~\ref{Fig1}(c).
Furthermore, the sign of the ratio \( t_2/t_1 \) controls the position of the NL within the Brillouin zone, shifting it between the zone center (\( \Gamma \) point) and the zone corner (M point), as detailed in Fig.~S2 of the SM. This minimal TB model thus explicitly demonstrates that type-II NLs arise from a delicate balance between intra- and interlayer electronic interactions, providing a general design principle for the targeted engineering of type-II NLSMs.

Building upon this theoretical foundation, we now translate these insights into a concrete, material-specific design strategy for realizing 2D type-II NLSMs, as schematically illustrated in Fig.~\ref{Fig1}(d). This bottom-up design strategy unfolds across three pivotal stages:
\begin{enumerate}
    \item \textbf{Parent Monolayer Selection:} The process begins with an isolated monolayer hosting a parabolic valence-band maximum (VBM) or conduction-band minimum (CBM) that is energetically well-separated from other bands. This ensures a simple, massive fermionic character as the foundational building block.
    \item \textbf{Mirror-Symmetric Bilayer Stacking:} Two identical monolayers are then stacked to form a bilayer structure with preserved vertical mirror symmetry ($\mathcal{M}_z$). This symmetry-guaranteed interlayer coupling splits the original band into a pair of mirrored counterparts. Crucially, their identical effective masses set the stage for the subsequent emergence of tilt-induced type-II crossings.
    \item \textbf{Intercalation-Mediated Fine-Tuning:} The final step involves the introduction of guest atoms into the van der Waals gap. This intercalation serves a dual purpose: it selectively enhances the interlayer hybridization strength, and simultaneously gates the Fermi level ($E_F$), thereby precisely locking the type-II NL at the chemical potential.
\end{enumerate}

A key advantage of this approach lies in its experimental feasibility. Steps (ii) and (iii) leverage well-established van der Waals stacking and intercalation techniques \cite{yu2021intercalation, zhou2021layered, li2024intercalation, yang2024intercalation, cao2021emerging, liu2025precision}, which have been successfully demonstrated in a wide range of 2D materials including bilayer graphene \cite{lin2024alkali, pakhira2018dirac, ma2025superconductivity}, $h$-BN \cite{radhakrishnan2017fluorinated, salpekar2024multifunctional, meiyazhagan2021gas, doping-BN, lonvcaric2018strong, shimada2017theoretical, sun2023regulating}, and MoS$_2$ \cite{wang2024achieving, wu2022observation, nong2025cu, wang2024ultrafast}. The proven capability to tune phenomena like superconductivity \cite{asaba2017rotational, wu2019spacing, zhang2020enhancement, zhang2022emergent, liang2023ni} or excitonic effects \cite{yao2014optical,bao2014approaching,zhang2022breaking} through these methods provides a solid foundation for implementing our band-structure design principle. This experimental readiness enables us to directly transition to material realization, beginning with the identification of suitable monolayer platforms.

\begin{figure}
	\includegraphics[width=3.5 in]{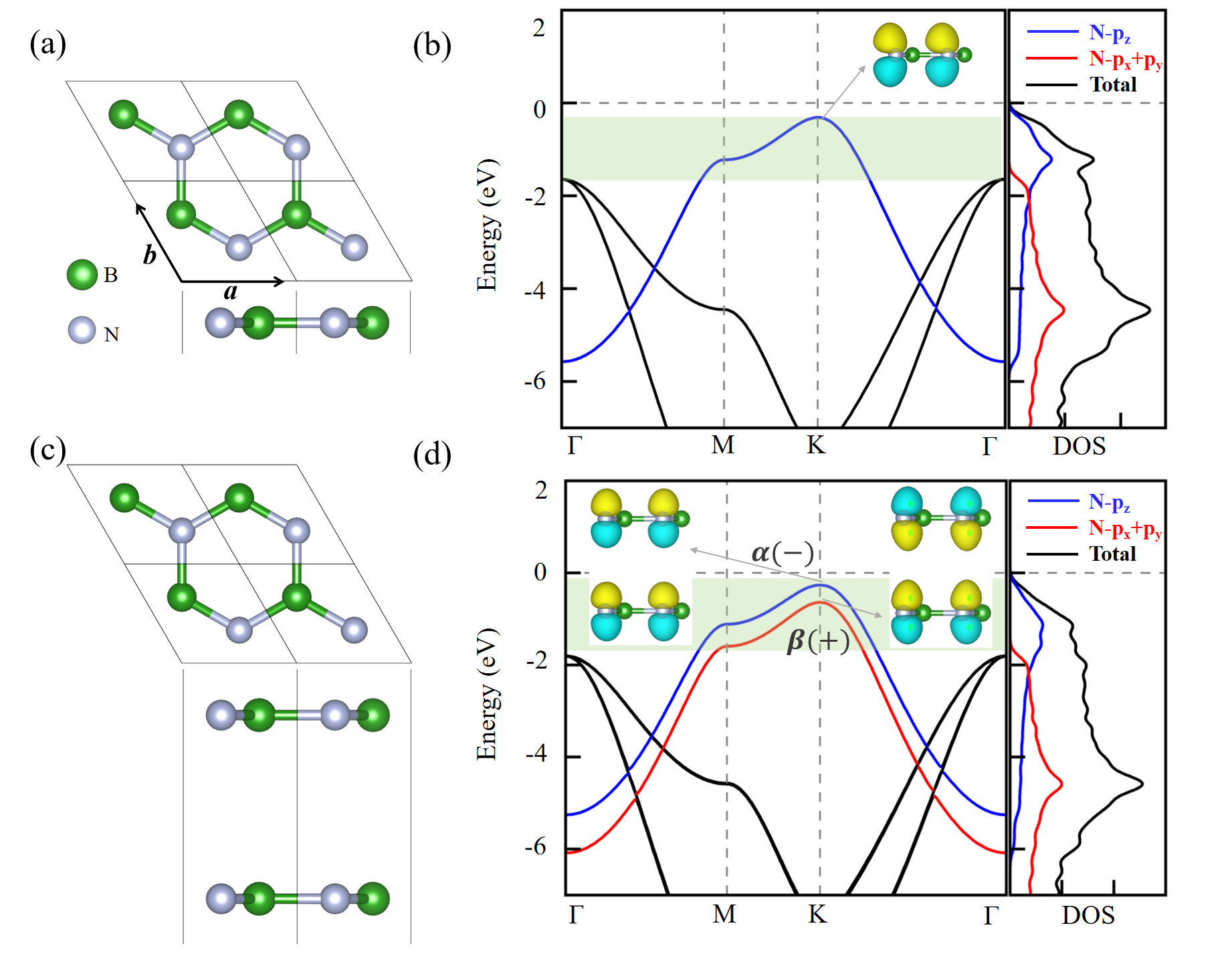}
	\caption{(a,c) Top and side views of the crystal structures of monolayer (ML) and AA-stacked bilayer (BL) $h$-BN, respectively.
    (b,d) Electronic band structures and projected density of states (PDOS) of ML and BL $h$-BN, respectively, with  
    with the wave-function distributions for the $\alpha$ and $\beta$ bands around the K point shown alongside.}\label{Fig2}
\end{figure}

\begin{figure*}
	\includegraphics[width=7 in]{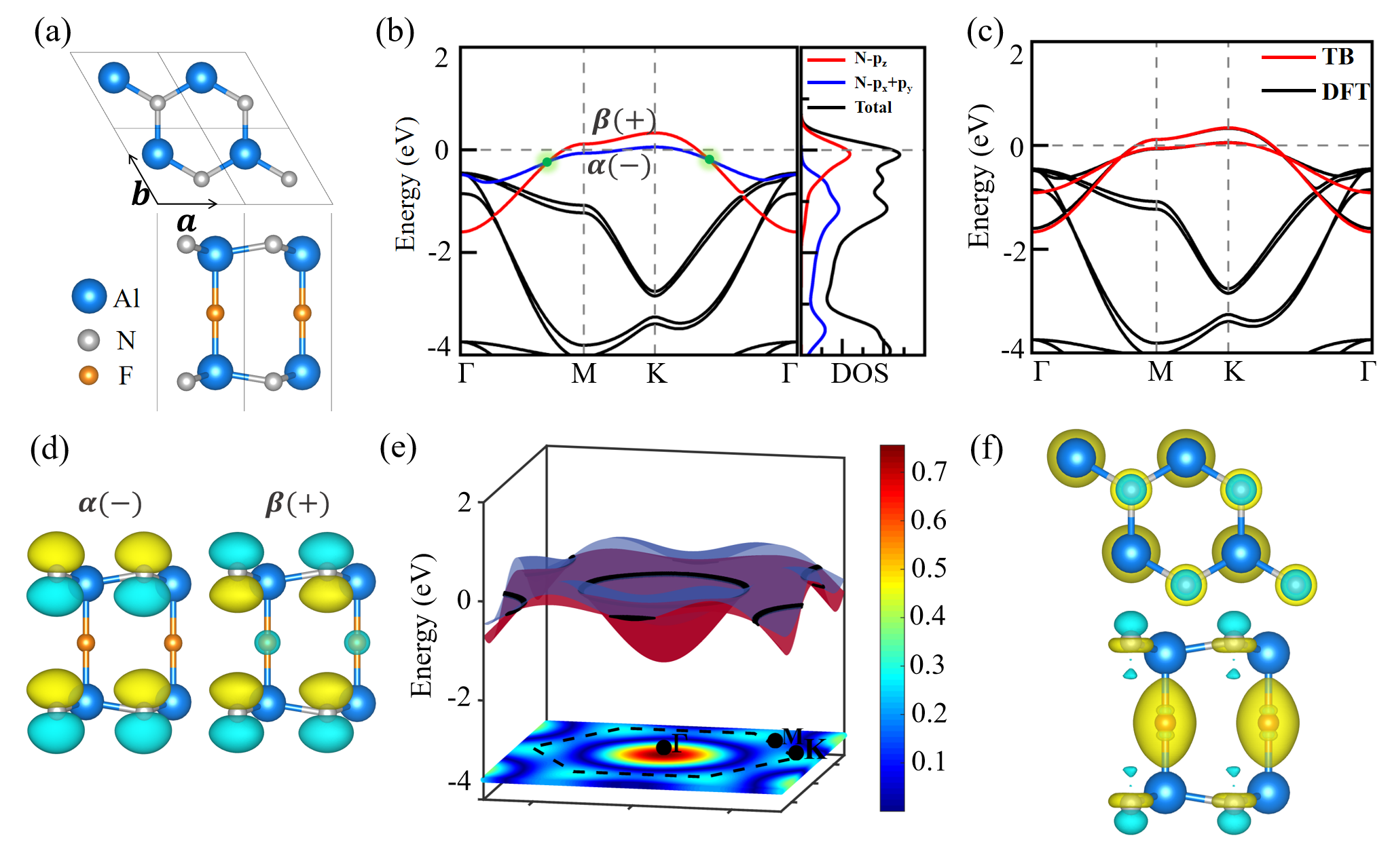}
\caption{(a) Top and side views of the atomic structure of fluorine-intercalated bilayer AlN (F@BL-AlN). (b) Electronic band structure and PDOS of F@BL-AlN. (c) Comparison between the TB and DFT band structures near the Fermi level. (d) Real-space distributions of the $\alpha$ and $\beta$ states at the K point. (e) Top and side views of the 3D band dispersion around the nodal line. (f) Differential charge density of F@BL-AlN, where yellow (blue) regions indicate charge accumulation (depletion).}\label{Fig3}
\end{figure*}

\section{Materialization of type-II NLSM in Hexagonal Nitrides}	
Based on the proposed design criteria, we systematically explore 2D monolayers that host atomically clean band edges as potential platforms for type-II nodal lines (NLs). Among the promising candidates, the family of hexagonal nitrides ($h$-XN, X = B, Al, Ga) stands out as an ideal material system [Fig.~S3 in SM], featuring isolated and parabolic band extrema---either valence band maxima (VBM) or conduction band minima (CBM)---that remain well-separated from other band manifolds \cite{wickramaratne2018monolayer,keccik2015layer}. We focus on monolayer $h$-BN as a prototypical system, chosen not only for its experimental accessibility, ideal band topology, and outstanding environmental stability, but also due to recent advances in stacking-engineered configurations---including moir\'e superlattices\cite{zhao2021universal,woods2021charge,wzz2-pszx}, sliding ferroelectricity\cite{yasuda2021stacking}, and tailored interlayer coupling in multilayer forms\cite{PhysRevB.93.245438,radhakrishnan2017fluorinated,salpekar2024multifunctional}. As shown in Fig.~\ref{Fig2}(a,b), the valence band maximum, dominated by N-$p_z$ orbitals, lies more than 1.5~eV above the next-higher valence band, cleanly satisfying the step (i) requirement of a pristine band edge.

Moving to step (ii), we construct mirror-symmetric bilayers by stacking two $h$-BN monolayers in the AA configuration while preserving $\mathcal{M}_z$ symmetry [Fig.~\ref{Fig2}(c)]. The interlayer coupling splits the VBM into two branches, $\alpha$ and $\beta$, characterized by opposite mirror eigenvalues but nearly identical effective masses [Fig.~\ref{Fig2}(d)]. However, the splitting remains positive throughout the Brillouin zone, preventing band inversion and consequently excluding nodal line formation. A similar behavior is observed in AA-stacked $h$-AlN and $h$-GaN bilayers (Fig.~S4). Our tight-binding analysis (Tab.~S1) reveals that this limitation stems from the dominance of vertical hopping over the skew component ($t_1 \gg t_2$), which suppresses the band inversion necessary for nodal line emergence. Even under substantial in-plane ($\pm6\%$) and out-of-plane ($\pm15\%$) strain (Fig.~S5), a finite local gap persists between the two bands throughout the Brillouin zone, confirming that pristine AA stacking alone cannot induce nodal lines and that step (ii) by itself remains insufficient.

To overcome these intrinsic constraints, we implement step (iii) via atomic intercalation as a targeted strategy to tailor interlayer coupling in BL-XN (X = B, Al, Ga) systems. We systematically evaluate seven second-period elements (Li--F) inserted at four high-symmetry sites (X-middle, N-middle, bridge, and hollow), resulting in 28 distinct configurations per BL-XN system (see SM Section~III). Through total-energy calculations, phonon dispersion analysis, and \textit{ab initio} molecular dynamics simulations, only two intercalated configurations emerge as thermodynamically and dynamically stable: F at the Al-middle site in BL-AlN (denoted F@BL-AlN) and F at the Ga-middle site in BL-GaN (F@BL-GaN). Although BL-BN intercalation exhibits intriguing electronic features, it remains dynamically unstable. We therefore select F@BL-AlN as a representative case to illustrate how atomic intercalation can transform a trivial bilayer insulator into a type-II NLSM.

Focusing on F@BL-AlN [Fig.~\ref{Fig3}(a)], first-principles calculations reveal that F atoms preferentially occupy Al-middle sites, forming a stable $P\bar{6}m2$ structure that maintains $\mathcal{M}_z$ symmetry. The resulting band structure [Fig.~\ref{Fig3}(b)] exhibits two branches, $\alpha$ and $\beta$, carrying opposite $\mathcal{M}_z$ eigenvalues and intersecting linearly near $E_F$ along the $\Gamma$–M and K–$\Gamma$ paths—clear evidence of interlayer-hybridization-driven band inversion. The three-dimensional dispersion in Fig.~\ref{Fig3}(e) demonstrates that these branches form a closed nodal loop around $\Gamma$, with each $k$-point exhibiting overtilted electron-hole mixed dispersion, unambiguously confirming F@BL-AlN as an overtilted type-II NLSM. Real-space wavefunction analysis reveals bonding–antibonding character between layers [Fig.~\ref{Fig3}(d)], similar to the pre-intercalation case, guaranteeing opposite mirror eigenvalues and enforcing symmetry-protected nodal crossings.

Strikingly, these electronic characteristics are accurately reproduced by a minimal TB model with parameters $e_0 = -0.3$~eV, $t = -0.165$~eV, $t_1 = 0.03$~eV, and $t_2 = 0.058$~eV. Based on our model analysis (Fig.~S1), the ratios $t_2 / t_1 = 1.933 > 1/2$ and $|t_2 / t| = 0.3515 < 1$ satisfy the criteria for forming a type-II nodal loop centered at $\Gamma$. This model successfully captures the interlayer-driven band inversion and the resulting type-II nodal loop, showing excellent agreement with DFT results [Fig.~\ref{Fig3}(c)]. Similarly, an ideal type-II NLSM state emerges near the Fermi level in F@BL-GaN (Fig.~S11), further affirming the generality of this intercalation-induced mechanism.

It is particularly noteworthy that fluorine intercalation in BL-AlN serves a dual function. Beyond subtly modifying interlayer couplings to fulfill the symmetry and band-inversion requirements for nodal line formation, it induces substantial electronic charge redistribution that shifts the Fermi level to the nodal energy. The differential charge-density map [Fig.~\ref{Fig3}(f)] reveals pronounced, directional charge transfer from N to F atoms, which partially depletes the N-derived $\alpha$ and $\beta$ bands and thereby aligns their inverted crossing precisely with $E_F$. Bader charge analysis quantitatively confirms this effect: each N atom gains ~2.0 $e$, each intercalated F atom accepts ~0.8 $e$, and the two Al atoms collectively contribute ~4.8 $e$. Thus, fluorine intercalation not only stabilizes the bilayer structure but also provides the essential charge imbalance, culminating in a tunable 2D type-II NLSM.

\begin{figure*}
		\includegraphics[width=7 in]{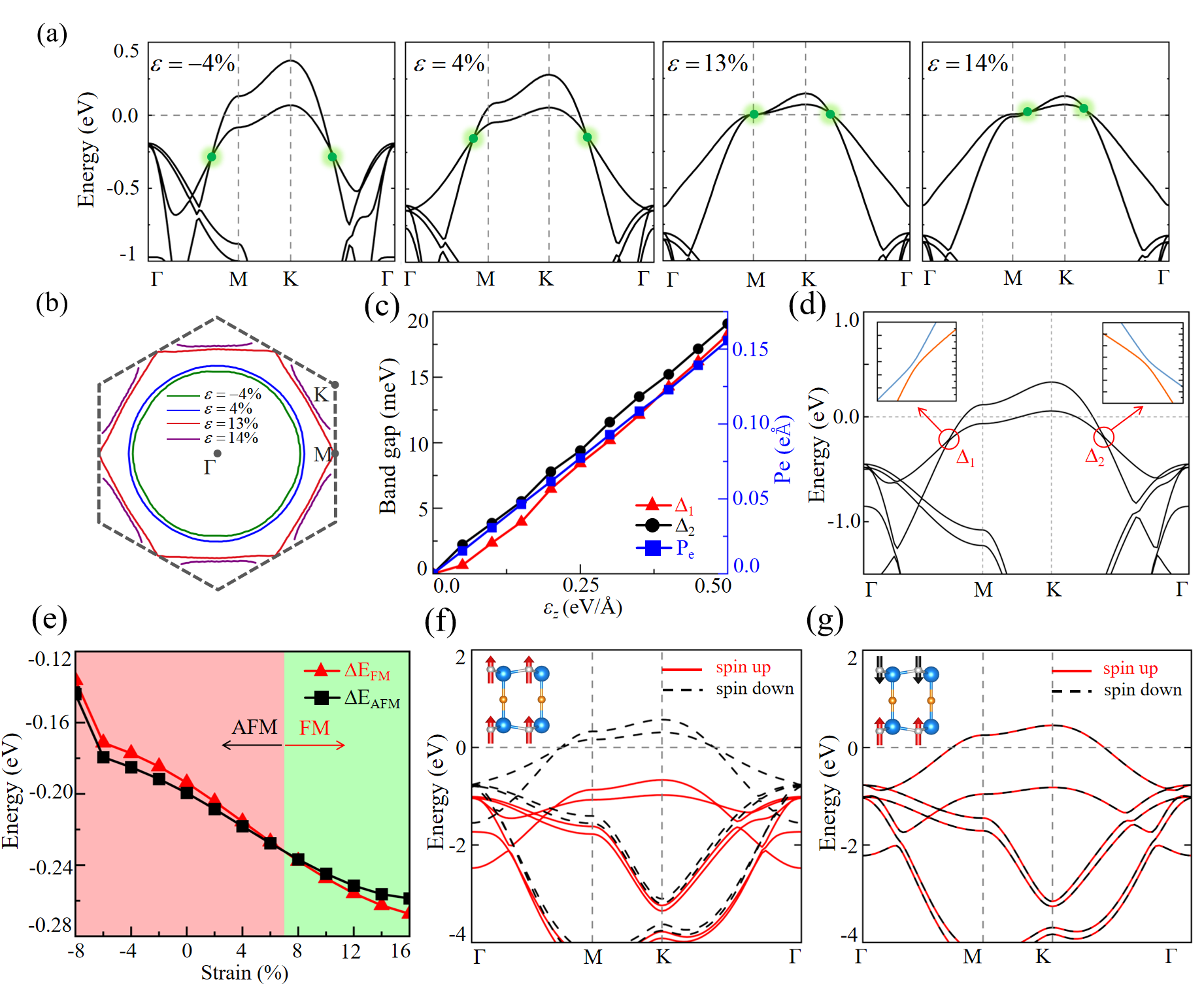}
		\caption{(a) Electronic band structures of F@BL-AlN under 
        biaxial strains. (b) Evolution of the NL structure with applied strain. (c) Evolution of the local band gaps ($\Delta_1$, $\Delta_2$) and electric polarization ($P_e$) under increasing vertical electric field,  $\mathcal{E}_z$. (d) Band structures of F@AlN with $\mathcal{E}_z$=0.25 eV/\AA. (e) Total energy of F@BL-AlN under FM and AFM configurations as a function of applied strain, with the nonmagnetic state taken as the zero-energy reference. (f) Spin-polarized band structure of BL-F@AlN in the FM configuration. (g) Spin-polarized band structure of BL-F@AlN in AFM.}\label{Fig4}
\end{figure*}

\section{Tunable physical properties of type-II NLSM in F@BL-AlN}	
The 2D type-II NLSM realized in F@BL-AlN arises from inverted $p_z$ bonding–antibonding states. Owing to their opposite eigenvalues under mirror symmetry $\mathcal{M}_z$, these states avoid hybridization, leading to a symmetry-protected nodal loop (NL) near the Fermi level ($E_F$). In analogy to three-dimensional NLSMs\cite{PhysRevLett.115.036806,PhysRevB.92.081201,PhysRevB.111.125101}, such band crossings are generally associated with a quantized Berry phase computed along a closed path encircling the NL. In two dimensions, however, no rigorous topological invariant uniquely defines a NL. Instead, a 2D NLSM is more appropriately viewed as a symmetry-protected semimetallic critical phase, serving as a parent state for various correlated insulating phases. For instance, introducing spin–orbit coupling (SOC) can drive a 2D NLSM into a quantum spin Hall insulator or a crystalline topological insulator characterized by a nontrivial mirror Chern number; alternatively, magnetic order may induce a transition to a Chern insulator\cite{PhysRevB.102.155111}. In the present F@BL-AlN system, however, SOC is negligible due to the light atomic masses of the constituent elements, as confirmed by our calculations (Fig.~S12). Hence, in the following, we explore the evolution of the electronic structure and topology under external perturbations, emphasizing symmetry breaking effects while justifiably neglecting SOC.

Under biaxial strain—which preserves $\mathcal{M}_z$—the nodal line remains gapless, yet undergoes a continuous geometric evolution. As illustrated in Figs.~\ref{Fig4}(a) and \ref{Fig4}(b), the band crossing points shift toward $E_F$ and the NL expands in momentum space. At a tensile strain of $13\%$, the NL touches the Brillouin zone boundary at the M point. Beyond this critical strain, the single large nodal ring fragments into smaller nodal structures centered around the two inequivalent K points, signifying a topological phase transition accompanied by a band inversion at M. This strain-driven restructuring of the nodal topology stems from the modulation of the hopping ratio $t_2/t_1$, in full agreement with our tight-binding model predictions [Fig.~S1(d)]. These results underscore the remarkable robustness of the type-II NLSM against symmetry-preserving mechanical deformations.

In contrast, breaking $\mathcal{M}_z$ offers a direct pathway to opening a band gap. As demonstrated in Figs.~\ref{Fig4}(c) and \ref{Fig4}(d), applying a perpendicular electric field $\mathcal{E}_z$ introduces a layer-staggered potential that explicitly breaks $\mathcal{M}_z$ and lifts the degeneracy between the $\alpha$ and $\beta$ states, thereby inducing local gaps along the original NL. The resulting gap magnitude exhibits an approximately linear dependence on $\mathcal{E}_z$, with coefficients of $\sim 0.037\ \mathrm{e\AA}$ (along $\Gamma$–M) and $\sim 0.038\ \mathrm{e\AA}$ (along K–$\Gamma$). Concurrently, the system develops an electronic polarization $P_e$ that partially screens the external field, enabling continuous and reversible electrical control over the band gap. For instance, at $\mathcal{E}_z = 0.25\ \mathrm{V/\AA}$, the gaps at high-symmetry points $\Lambda_1$ and $\Lambda_2$ reach $\sim 6.5$ meV and $7.8$ meV, respectively.

Unlike type-I NLSMs that typically exhibit a suppressed density of states (DOS) near the nodal energy, the type-II NLSM in F@BL-AlN features a pronounced DOS peak at $E_F$ [Fig.~\ref{Fig3}(b) and Fig.~S11]. This enhanced spectral weight renders the system highly prone to interaction-driven instabilities upon breaking time-reversal symmetry ($\mathcal{T}$). Indeed, spin-polarized total-energy calculations reveal that both F@BL-AlN and F@BL-GaN spontaneously develop long-range magnetic order. The magnetic behavior originates from F-induced charge redistribution, which localizes an unpaired electron on each N atom and creates local moments of approximately $0.5~\mu_B$ per site. The substantial DOS at $E_F$ satisfies the Stoner criterion, supporting the emergence of itinerant $p$-electron magnetism\cite{yi2025two,PhysRevLett.102.017201,fu2017effects}. Comparative total-energy evaluations across various magnetic configurations [Fig.~S13 and Tab.~S5] indicate that F@BL-AlN adopts an intralayer ferromagnetic and interlayer antiferromagnetic (AFM) ground state, with a predicted Néel temperature ($T_N$) of 156 K [Fig.~S14(a)]. In contrast, F@BL-GaN stabilizes in a fully ferromagnetic (FM) state, exhibiting a Curie temperature ($T_C$) of 150 K [Fig.~S14(b)]\cite{LIU2019300,liu2020symmetry}. Furthermore, biaxial strain effectively modulates the magnetic energetics [Fig.~\ref{Fig4}(e)]. In F@BL-AlN, the AFM phase remains favorable under moderate strain, yet substantial tensile strain ($\gtrsim 7\%$) destabilizes antiferromagnetism and triggers an AFM–FM transition, indicative of a strain-mediated magnetic phase reversal. Conversely, F@BL-GaN maintains a robust FM ground state across a wide strain range, transitioning to the AFM phase only under strong compression ($\lesssim 8\%$) [Tab.~S6].

Magnetism, in turn, selectively reshapes the band topology in a symmetry-dependent manner. In the FM phase, exchange splitting polarizes the two spin channels, yet nodal features persist within each spin sector due to the underlying $\mathcal{M}_z$-governed band connectivity, yielding a fully spin-polarized type-II NLSM. In the collinear AFM ground state of F@BL-AlN [Fig.~\ref{Fig4}(g)], $\mathcal{M}_z$ is broken, but the combined anti-unitary symmetry $\mathcal{M}_z\mathcal{T}$ is preserved. This symmetry enforces a twofold degeneracy at every $\mathbf{k}$-point, placing the system within the novel class of type-IV antiferromagnets\cite{tian2025symmetry, bai2025anomalous}. Such materials are characterized by Kramers-like degeneracy in the absence of spatial inversion—a distinctive signature amenable to experimental detection. These results demonstrate that strain provides a practical route for the concurrent control of topology and magnetism in F@BL-AlN via a single mechanical parameter.

\section{Discussion and conclusion}

In summary, we present a general bottom-up strategy for realizing 2D type-II NLSMs by harnessing the combined power of van der Waals stacking and atomic intercalation. Using the well-established \(h\)-BN family as a representative platform, we identify F@BL-AlN and F@BL-GaN as ideal 2D type-II NLSMs, where mirror symmetry protects the nodal loops and the negligible spin–orbit coupling preserves their pristine character. These nodal-line states are highly responsive to external tuning: biaxial strain reshapes the nodal geometry and drives a symmetry-preserving topological phase transition, while electric fields or structural asymmetry lift the mirror protection and open controllable band gaps. 
Notably, in contrast to type-I NLSMs, the large density of states intrinsic to the type-II dispersion amplifies electronic correlations, stabilizing an $\mathcal{M}_z\mathcal{T}$ -protected antiferromagnetic ground state or enabling spin-polarized nodal features in the ferromagnetic phase. Together, these results establish F@BL-AlN and F@BL-GaN as versatile platforms where topology and magnetism naturally intertwine.
These findings collectively illustrate that the intercalation-based stacking approach may provide a versatile materials platform for engineering tunable type-II topological states.

\begin{acknowledgements}
This work was supported by the Natural Science Foundation of China (Grant No. 12204330). B. Fu acknowledges financial support from Sichuan Normal University (Grant No. 341829001). The numerical computations were conducted at the Hefei Advanced Computing Center, with additional support from the High-Performance Computing Center of Sichuan Normal University, China.

\end{acknowledgements}

\bibliographystyle{apsrev4-2}

\end{document}